\documentclass{article}
\usepackage{ijcai22}

\usepackage{times}
\usepackage{soul}
\usepackage{url}
\usepackage[hidelinks]{hyperref}
\usepackage[utf8]{inputenc}
\usepackage[small]{caption}
\usepackage{graphicx}
\usepackage{amsmath}
\usepackage{amsthm}
\usepackage{booktabs}
\usepackage{algorithm}
\usepackage{algorithmic}
\usepackage{xcolor}
\usepackage{multirow}

\title{A Physics-Informed Neural Network to Model Port Channels}

\author{
Marlon S. Mathias\footnote{Main and corresponding author.
All other authors made significant contributions and the last five authors
oversaw the whole project and secured funding.}$^1$
\and
Marcel R. de Barros$^2$\and
Jefferson F. Coelho$^2$\and
Lucas P. de Freitas$^2$\and\\
Felipe M. Moreno$^2$\and
Caio F. D. Netto$^2$\and
Fabio G. Cozman$^2$\and \\
Anna H. R. Costa$^2$\and
Eduardo A. Tannuri$^2$\and
Edson S. Gomi$^2$\and
Marcelo Dottori$^3$\and
\affiliations
Center for Artificial Intelligence (C4AI) -- University of Sao Paulo, Brazil\\
$^1$Instituto de Estudos Avançados -- University of São Paulo, Brazil\\
$^2$Escola Politécnica -- University of Sao Paulo, Brazil\\
$^3$Instituto Oceanográfico -- University of Sao Paulo, Brazil\\
\emails
\{marlon.mathias, caio.netto, marcel.barros, jfialho, lfreitasp2001, felipe.marino.moreno, fgcozman, \\ anna.reali, eduat, gomi, mdottori\}@usp.br
}

\begin{document}

\maketitle

\begin{abstract}
    We describe a Physics-Informed Neural Network (PINN) that simulates the flow induced by the astronomical tide in a synthetic port channel, with dimensions based on the Santos -- São Vicente -- Bertioga Estuarine System. PINN models aim to combine the knowledge of physical systems and data-driven machine learning models. This is done by training a neural network to minimize the residuals of the governing equations in sample points. In this work, our flow is governed by the Navier-Stokes equations with some approximations. There are two main novelties in this paper. First, we design our model to assume that the flow is periodic in time, which is not feasible in conventional simulation methods. Second, we evaluate the benefit of resampling the function evaluation points during training, which has a near zero computational cost and has been verified to improve the final model, especially for small batch sizes. Finally, we discuss some limitations of the approximations used in the Navier-Stokes equations regarding the modeling of turbulence and how it interacts with PINNs.
\end{abstract}

\section{Introduction}



\begin{figure*}[t]
    \centering
    \includegraphics[width=0.9\textwidth]{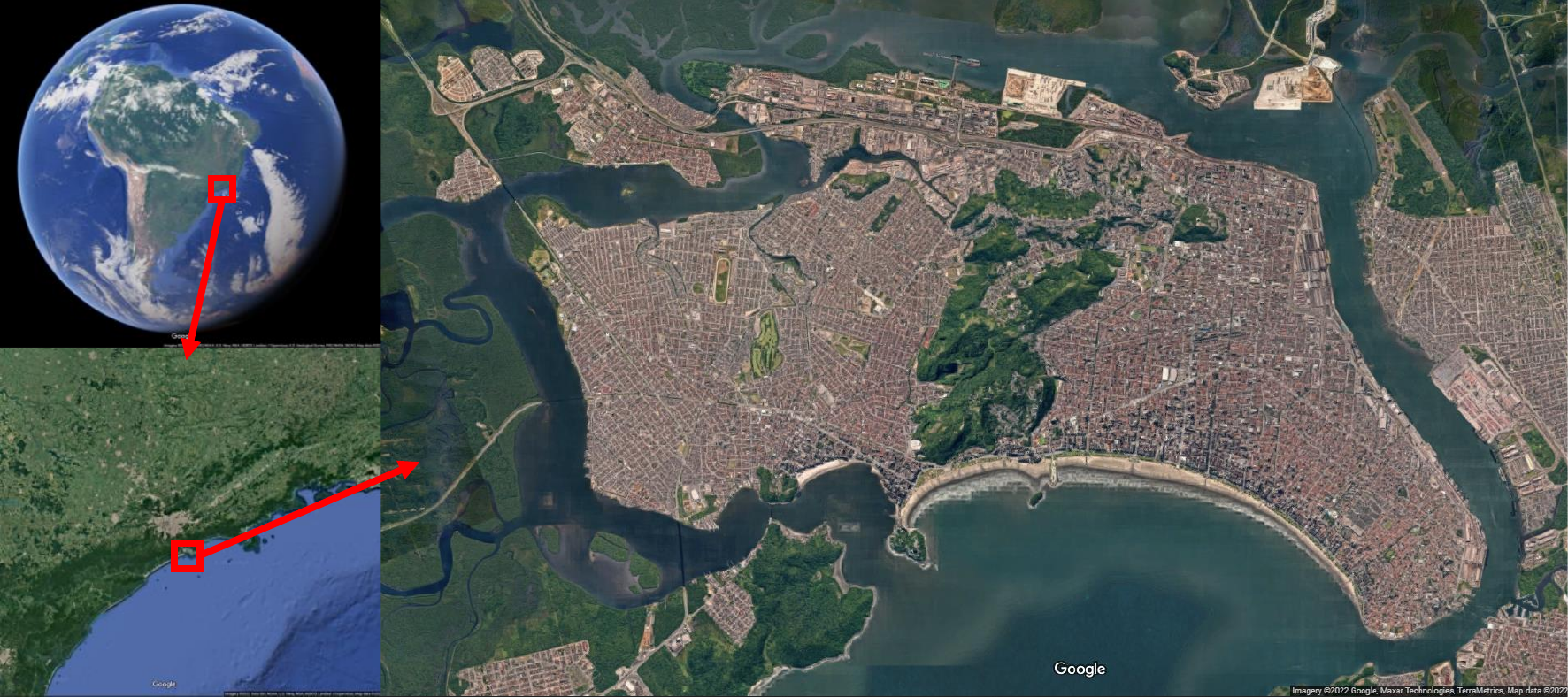}
    \caption{Satellite image of the Santos -- S\~ao Vicente -- Bertioga Estuarine System. (Adapted from Google\textsuperscript{\textregistered} Maps, \url{https://maps.google.com}).}
    \label{fig:map}
\end{figure*}

The Santos -- S\~ao Vicente -- Bertioga Estuarine System houses the Port of Santos, which is of fundamental importance for Brazilian trade and economy. Therefore, it is of utmost importance to have a reliable system to model and forecast oceanographic variables in that region. Figure~\ref{fig:map} shows satellite images of the estuarine system.

\cite{costa2020} have developed the Santos Operational Forecasting System (SOFS), which monitors and predicts variables such as tides, currents, salinity, and temperature near the port. This model is based on the Princeton Oceanic Model (POM-rain), by \cite{blumberg1987}, which numerically approximates the solution of the Navier-Stokes equations that describes fluid flows.

Fluid simulation is notoriously complex and expensive, and consequently some approximations must be made for them  to become computationally feasible, with a balance between accuracy and computational cost. The chaotic nature of fluids adds to this challenge, as small errors or short-term approximations largely affect the accuracy of predictions if a long forecast horizon is required.

The rapid development of Machine Learning (ML) models in recent years has led to great progress in various areas and allowed a wholly new approach to problems involving time-series forecasts. By training these models with large datasets, the models can recognize complex patterns and address the prediction  of   physical system. However, a purely data-driven approach to the problem of ocean forecasting is also not completely desirable, as such models require very large amounts of data to train, which might not always be available, and they may lead to non-physical solutions when presented to previously unseen scenarios \cite{karniadakis2021}.


Physics-Informed Machine Learning models present a compromise between the two approaches in an attempt to combine their advantages and minimize their shortcomings. The literature presents several different ways of combining physical and data-driven models \cite{kashinath2021,raissi2019,willard2020}. These approaches include, but are not limited to: using ML to estimate the error in the physics-based models \cite{xu2015}; using ML to increase the resolution of known flow fields \cite{nair2019,fukami2020}; substituting the governing equations by trained neural networks \cite{wu2020}; adding physical constraints to the ML model \cite{beucler2021,read2019}.

In this work  we propose a Physics-Informed Neural Network (PINN) that uses the governing equations of the physical system in its fitness evaluation. The solution of the equation is unknown in the whole domain, and the ML model is used to approximate it. \cite{li2021} calls this approach a neural-FEM, comparing it to the Finite Elements Method (FEM) of solving partial differential equations. In this analogy, the neural network works as one large and complex finite element, which spans the whole domain and solves the equation within its solution space, similarly to the classical FEM, in which several elements, each with a relatively simple solution space, are spread along the domain in a mesh.

\begin{figure}[t]
    \includegraphics[width=0.5\textwidth]{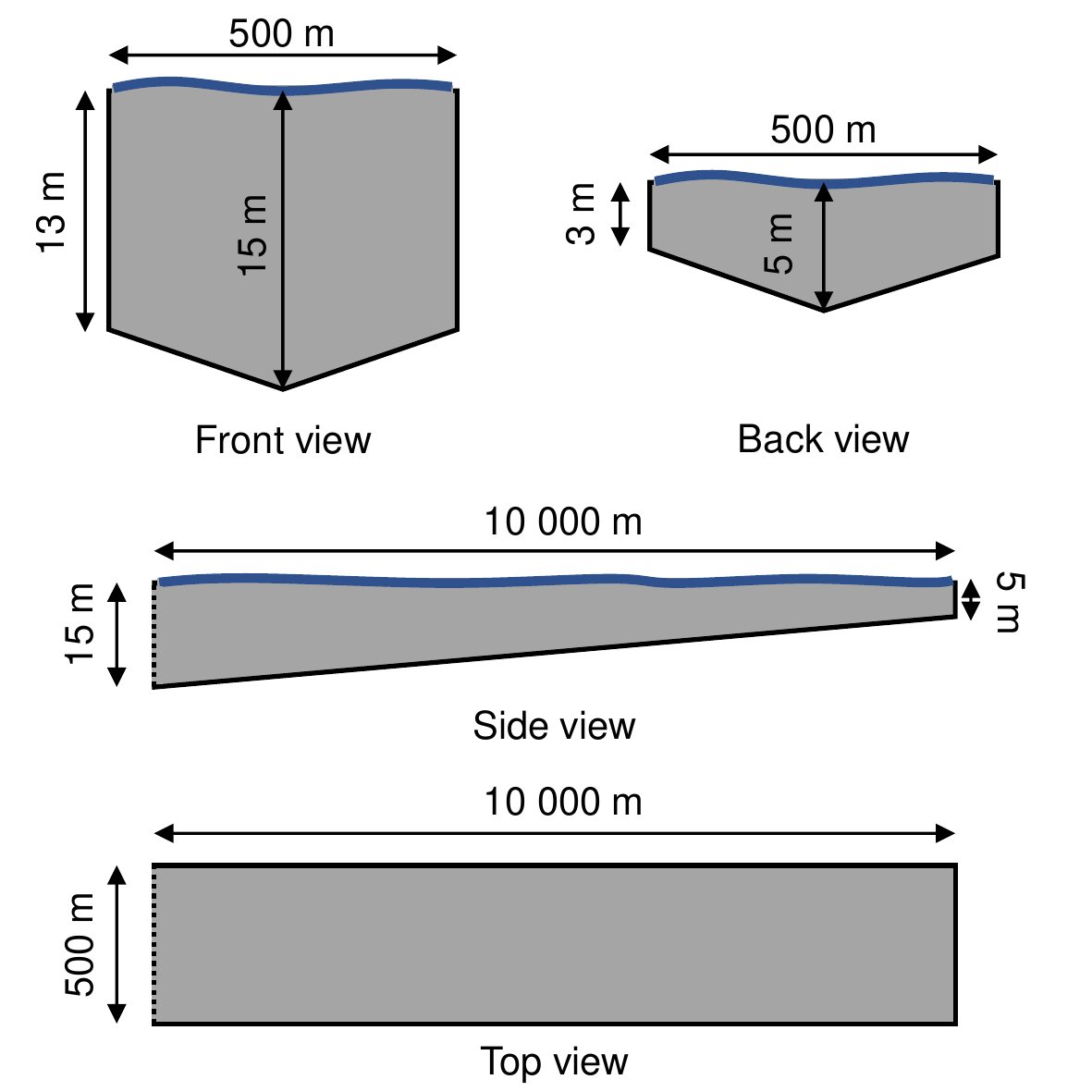}
    \caption{Channel dimensions (not to scale). The blue, wavy line denotes the free surface of the water, and the dotted line denotes the open boundary of the channel.}
    \label{fig:dimensions}
\end{figure}

We focus on a synthetic, rectangular port channel whose dimensions
 resemble those of the Port of Santos.
It is 10 km long and 500 m wide. At the entrance, the depth is 15 m in the center and 13 m at both sides, at its other end, the depths are linearly reduced to 5 m and 3 m, respectively. Figure~\ref{fig:dimensions} depicts the channel. Note that the drawings are not to scale due to the different orders of magnitude in the lengths. The open end of the channel is exposed to the astronomical tide and its 12-hour cycle. All other sides are closed. 

This paper makes two contributions: First, we demonstrate the capability of a PINN model to reach the fully developed solution of a system that is periodic in time while skipping the transient phase, as opposed to more traditional ways of simulation, which need a long accommodation phase, when the effects of the initial condition are still present.
The other contribution of this work is to evaluate the effect of resampling the function evaluation points during the training of the PINN. This has allowed us to maintain the error margins while reducing the number of function points evaluated at once. This smaller batch size, in turn, reduces the computational cost, especially in terms of the memory footprint, allowing either cheaper computers to train the model or larger models to be trained in a given computer.

The remainder of this paper is structured as follows: In Section~\ref{sec:equations}, we present the governing equations of our physical system, its boundary conditions and physical assumptions that were made. In Section~\ref{sec:pinn}, we describe the PINN implementation and training procedure. Section~\ref{sec:results} shows the results obtained after training was complete. Finally, Section~\ref{sec:conclusion} finishes with some conclusions and suggestions for future works.

\section{Governing Equations and Boundary Conditions}
\label{sec:equations}


In our systems of interest, flow is governed by the Navier Stokes equations, as described by \cite{blumberg1987} and implemented in their POM-rain scheme. A mode splitting technique is used to separate the three dimensions in an external mode, which comprises the horizontal directions ($x$ and $y$), and an internal mode, which corresponds to the vertical direction ($z$); this way the number of variables to be solved at once is greatly reduced. Moreover, the cited authors suggest that the internal mode can be removed altogether, to further simplify the problem. Here we make the following approximations:
\begin{enumerate}
	\item Removal the internal mode;
	\item Disregard Coriolis acceleration;
	\item Disregard viscous forces at the surface and at the bottom;
	\item The fluid is incompressible and has a constant density;
	\item Disregard convective terms.
\end{enumerate}

The mass conservation equation is then:
\begin{equation}
	\frac{\partial \eta}{\partial t} + \frac{\partial (\bar{U} D)}{\partial x} + \frac{\partial (\bar{V} D)}{\partial y} = 0 ,
	\label{eq:ns1}
\end{equation}
where $\eta$ is the water surface elevation with respect to some standard height, $\bar{U}$ and $\bar{V}$ are the flow velocities in $x$ and $y$ direction respectively -- the bar denotes that these velocities are averaged along the $z$ direction. In addition, $D$ is the total height of the water column, given by:
\begin{equation}
	D(t,x,y) = H(x,y) + \eta(t,x,y),
\end{equation}
 where $H$ is the depth at a given coordinate $(x,y)$.

The equation for momentum conservation in the $x$ direction becomes:
\begin{equation}
	\frac{\partial (\bar{U} D)}{\partial t} - \tilde{F}_x + gD\frac{\partial\eta}{\partial x} = 0,
	\label{eq:ns2}
\end{equation}

\noindent $g$ is the local acceleration of gravity and $\tilde{F}_x$ is a dissipative term, given by:

\begin{equation}
	\tilde{F}_x = \frac{\partial}{\partial x} \left[2 H \bar{A}_M \frac{\partial \bar{U}}{\partial x} \right] + \left[ H \bar{A}_M \left(\frac{\partial \bar{U}}{\partial y} + \frac{\partial \bar{V}}{\partial x} \right) \right] ,
	\label{eq:fx}
\end{equation}

\noindent where $A_M$ is the Smagorinsky diffusivity:
\begin{equation}
	\resizebox{.91\linewidth}{!}{$
		\displaystyle
		A_M = C \Delta x \Delta y \frac{1}{2} \left[ \left( \frac{\partial \bar{U}}{\partial x} \right)^2 + \frac{1}{2}\left( \frac{\partial \bar{U}}{\partial y} + \frac{\partial \bar{V}}{\partial x} \right)^2 + \left( \frac{\partial \bar{V}}{\partial y} \right)^2 \right] .
		$}
		\label{eq:am}
\end{equation}%

The equation for the momentum conservation in $y$ is analogous. The symbol $C$ denotes a non-dimensional parameter, which is set to 0.2 in this work. Symbols $\Delta x$ and $\Delta y$ denote the mesh spacings at each point of the grid. The model needs these values to estimate the effect of small eddies on the overall dissipation, as will be discussed later in this section. However, the method described in this work is meshless and, instead of a regular grid, uses a set of randomly distributed points in the domain. Therefore, we set $\Delta x$ and $\Delta y$ to the mean spacing in $x$ and $y$ in the domain. With a total length $L_{x_i}$ in the domain direction $x_i$, $n$ points to be uniformly distributed and considering there are 3 dimensions in the solution space, these parameters can be calculated by:

\begin{equation}
	\Delta {x_i} = \frac{L_{x_i}}{n^{1/3}}.
	\label{eq:deltax}
\end{equation}

We follow POM-rain's implementation and use a form of fluid modeling that is known as Large Eddy Simulation (LES). This type of fluid model only requires larger structures -- or eddies -- to be resolved by the solver. Smaller structures, which are too small for the simulation to capture, have their effect modeled by terms that are added to the governing equations \cite{zhiyin2015}. In turbulent flows, energy usually flows from large to small structures in the phenomenon known as Kolmogorov energy cascade \cite{kolmogorov1991}. In turn, small eddies cause a dissipative effect on the larger scales, which is usually modeled as a ``turbulent viscosity" term that is added to the fluid's own viscosity. For practical purposes, this type of modeling allows coarser grids to be used for simulations, drastically reducing their computational cost. If no form of turbulence model is used, the grid would need to be fine enough to capture all eddies in the fluid, down to the Kolmogorov scale, which describes the smallest eddies possible before the turbulent kinetic energy is dissipated by the fluid's viscosity, in what is known as Direct Numerical Simulation (DNS).

We now discuss the boundary conditions employed in modeling.

The synthetic channel is represented by a rectangular domain, with different boundary conditions in each side. We use the fact that the channel is symmetric to our advantage, imposing a symmetry boundary condition at its center line ($y=0$). Therefore, only the $y\ge 0$ region is considered. This condition is obtained by imposing a Neumann boundary condition for the elevation and the longitudinal velocity, such that their derivatives normal to the center line are null. A Dirichlet boundary condition is used for the transversal velocity, such that it is null at the center line. The inlet (at $x=0$) is exposed to the astronomical tide, therefore we use a Dirichlet boundary condition to set the elevation level to the sine wave,
\begin{equation}
	\eta(t,x=0,y) = A \sin(\omega t) ,
\end{equation}
\noindent with $A$ set to 1 meter and $\omega$ such that the period is 12 hours. Both components of the velocity are under a Neumann condition such that their derivative normal to the boundary is null. At the end and at the side of the channel ($x=x_f$ and $y=y_f$), we have set no-slip wall conditions, which impose zero velocity and a null derivative of elevation normal to the boundary. The Neumann boundary conditions are needed to make sure that the problem is well-posed.

Time is taken to be periodical, with a cycle length of 12 hours. Hence there is no need to set an initial condition. This is done as part of the Fourier Features that are encoded in the network. More details on their implementation are given in the following sections.

\section{Physics-Informed Neural Network}
\label{sec:pinn}

Our goal is to build a model that receives coordinates $t$, $x$ and $y$ and outputs the variables $\eta$, $\bar{U}$ and $\bar{V}$ for any point of a channel at any given time during the astronomical tide's 12-hour cycle. For this, we use a PINN model consisting of the following elements:

\begin{enumerate}
	\item Normalization layer for spatial coordinates;
	\item Fourier features computation for time;
	\item Multilayer Perceptron (MLP);
	\item Boundary conditions encoding.
\end{enumerate}

The normalization layer is used so that the spatial input coordinates ($x$ and $y$) are linearly transformed to a range between zero and one: 
\begin{equation}
	 \hat{x}_i = \frac{x_i}{L_{x_i}},
\end{equation}

\noindent where the $\hat{x_i}$ denotes the normalized $x_i$ coordinate.

The time input ($t$) receives a different treatment to reflect the temporal periodicity of the system. A Fourier Features layer is used, converting the value of $t$ to the pair $[\cos(\omega t), \sin(\omega t)]^T$, with $\omega$ chosen so that the period equals 12 hours.

The MLP network is fully connected and has 4 hidden layers of 20 neurons each, with a hyperbolic tangent activation function. Larger amounts of layers and neurons were tested with little impact on the results. Residual connections are added every two layers, which allows for better computations of the gradients and aids the convergence during training \cite{he2016}. The inputs come from both the normalization and the Fourier Features layers. There are three outputs: $\tilde{\eta}$, $\tilde{U}$ and $\tilde{V}$, one for each variable in the problem. The tilde denotes that these variables are the direct output of the MLP and may not observe the boundary conditions of the problem.

After the MLP, there is a Boundary-encoded output layer, as described by \cite{sun2020}. This layer makes sure that the Dirichlet boundary conditions are always observed. The layer works by combining the output of the MLP to a known function that perfectly observes the Dirichlet boundary condition in the following manner: 
\begin{equation}
	\eta = d_\eta(x,y) \tilde{\eta} + \eta_p,
\end{equation} 
\noindent where $\eta_p$ is the particular solution for the water elevation and $d_\eta$ is a distance function, which is continuous and differentiable, and is equal to zero where the value of $\eta$ is set by a Dirichlet boundary condition and non-zero everywhere else. The concept behind this definition is that the value of $\eta$ is fixed to $\eta_p$ at the boundary conditions but remains under control of the neural network everywhere else in the domain. This operation is analogous to $\bar{U}$ and $\bar{V}$. We have defined the following equations for $d$:
\begin{equation}
	\resizebox{.89\linewidth}{!}{$
		\displaystyle
		\begin{bmatrix}
			d_\eta \\
			d_{\bar{U}} \\
			d_{\bar{V}}
		\end{bmatrix}
		=
		\begin{bmatrix}
			\tanh(\alpha_b \hat{x}) \\
			\tanh(\alpha_b (1-\hat{x})) \tanh(\alpha_b (1-\hat{y})) \\
			\tanh(\alpha_b \hat{y}) \tanh(\alpha_b (1-\hat{x})) \tanh(\alpha_b(1-\hat{y}))
		\end{bmatrix}.
		$}
\end{equation}

The hyperbolic tangent function is used so that the value of $d$ is constant and close to one when away from the boundaries. $\alpha_b$ is set to 10.56, which causes $d$ to reach 0.99 when its distance to the boundary is equivalent to 25\% of the domain. The Neumann boundary conditions cannot be set by using this method and are implemented via an additional loss function during training, as will be shown later in this section.

The particular solution for this case is set as $\bar{U}_p = \bar{V}_p = 0$ and
\begin{equation}
	\eta_p = A \sin(\omega t) ,
\end{equation}

\noindent with $A$ set to 1 meter and $\omega$ such that the period is 12 hours. This solution observes all Dirichlet boundary conditions.

\begin{figure*}[!ht]
    \centering
    \includegraphics[width=0.8\textwidth]{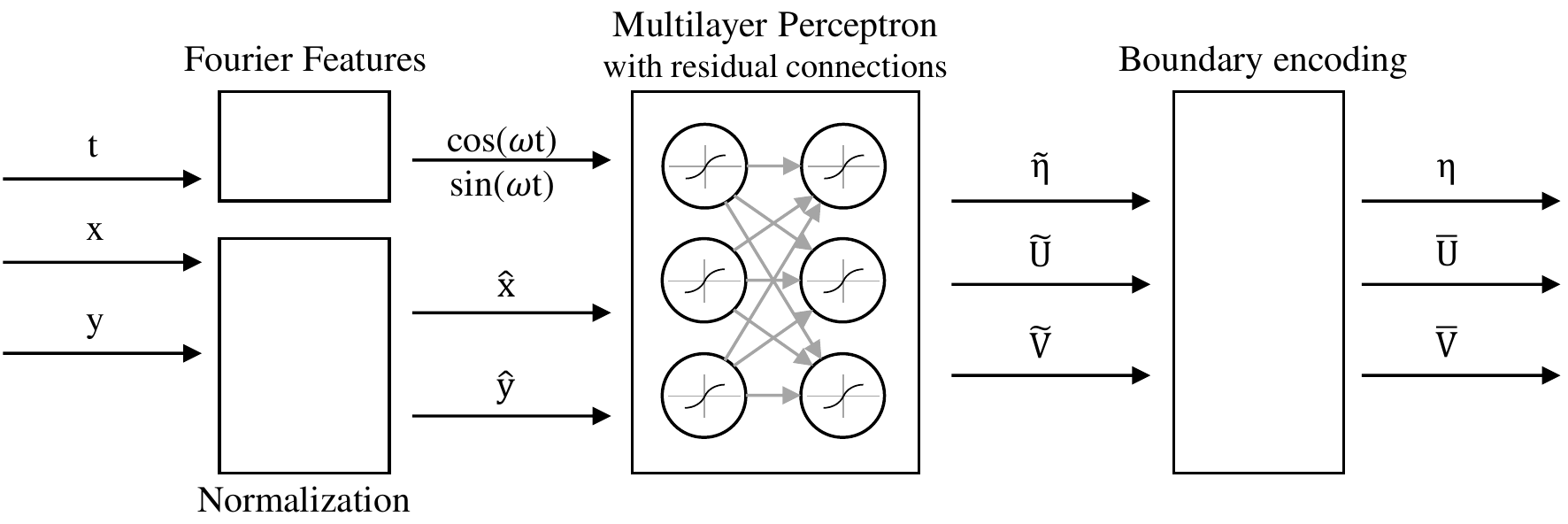}
    \caption{Summary of the PINN network.}
    \label{fig:network}
\end{figure*}

As a summary, Figure~\ref{fig:network} shows the final configuration of the network.

\subsection{Loss function}

We define the loss function of the network as the sum of the mean absolute value of the residual of each governing equation: mass conservation (Eq. \ref{eq:ns1}), and momentum conservation for both directions (Eq. \ref{eq:ns2}); for normalization purposes, each residual is multiplied by the total length in time of the domain ($L_t$). We also add to this loss function the mean absolute value of the Neumann boundary conditions residual, each multiplied by the total length of the domain in their respective direction ($L_x$ and $L_y$).

Therefore, given that $\eta_i$, $\bar{U}_i$ and $\bar{V}_i$ are the PINN's output at coordinate ($t_i$,$x_i$,$y_i$), the residuals relative to the governing equations are computed as:

\begin{equation}
	R_{f,1,i} = \left| \frac{\partial \eta_i}{\partial t} + \frac{\partial (\bar{U}_i D_i)}{\partial x} + \frac{\partial (\bar{V} D_i)}{\partial y} \right| ,
	\label{eq:r1}
\end{equation}

\begin{equation}
	R_{f,2,i} = \left| \frac{\partial (\bar{U}_i D_i)}{\partial t} - \tilde{F}_x + gD_i\frac{\partial\eta_i}{\partial x} \right| ,
	\label{eq:r2}
\end{equation}

\begin{equation}
	R_{f,3,i} = \left| \frac{\partial (\bar{V}_i D_i)}{\partial t} - \tilde{F}_y + gD_i\frac{\partial\eta_i}{\partial y} \right| ,
	\label{eq:r3}
\end{equation}

\noindent with $\tilde{F}_x$ computed by Eq.~\ref{eq:fx} and $\tilde{F}_y$ by its analogous in $y$. The residuals relative to the Neumann boundary conditions are given by:

\begin{equation}
    R_{b,1,i} = M_{\eta,i,x} \left| \frac{\partial \eta_i}{\partial x} \right| + M_{\bar{U},i,x} \left| \frac{\partial \bar{U}_i}{\partial x} \right| + M_{\bar{V},i,x} \left| \frac{\partial \bar{V}_i}{\partial x} \right| ,
\end{equation}

\begin{equation}
    R_{b,2,i} = M_{\eta,i,y} \left| \frac{\partial \eta_i}{\partial y} \right| + M_{\bar{U},i,y} \left| \frac{\partial \bar{U}_i}{\partial y} \right| + M_{\bar{V},i,y} \left| \frac{\partial \bar{V}_i}{\partial y} \right| .
\end{equation}
Note that $M_{Y,i}$ is either zero or one, depending on whether point $i$ has a Neumann condition for variable $Y$ in that particular direction.

The loss function of the model is given by:

\begin{equation}
    \resizebox{.89\linewidth}{!}{$
	\displaystyle
    \mathcal{L} = \frac{1}{n_f}\sum_{i=1}^{n_f}\sum_{n=1}^{3} L_t R_{f,n,i} + \frac{1}{n_b}\sum_{i=1}^{n_b}{\left(L_x R_{bx,i} + L_y R_{by,i}\right)}.
    $}
\end{equation}

The governing equations are evaluated at $n_f$ random points uniformly distributed along the domain, while the Neumann boundary conditions are evaluated at $n_b$ points uniformly distributed at the boundaries. One characteristic of PINNs is that there is an infinite pool of points for the loss to be evaluated. Consequently, the set of evaluation points can be changed as many times as desired. In this work, we compare models that were trained with a fixed set of points to ones trained with a set that is reshuffled between each iteration.

\subsection{Implementation}

Our code is written in Python and uses the PyTorch module for the machine learning tasks. The optimization is performed with the Adam algorithm. All tests were executed on a Nvidia\textsuperscript{\textregistered} GeForce\texttrademark~RTX3080 graphics card, with 10 GB of available memory. The code is available on GitHub\footnote{\url{https://github.com/marlonsmathias/PINN_Port_Channel}}.

\section{Results}
\label{sec:results}

The model was trained for various numbers of function evaluations, from $n_f=300$ to $n_f=100000$, both reshuffling the training batch every iteration and keeping it fixed. The number of boundary conditions evaluation points was kept fixed at $n_b=1000$.

\begin{figure}[!ht]
    \centering
    \includegraphics[width=0.45\textwidth]{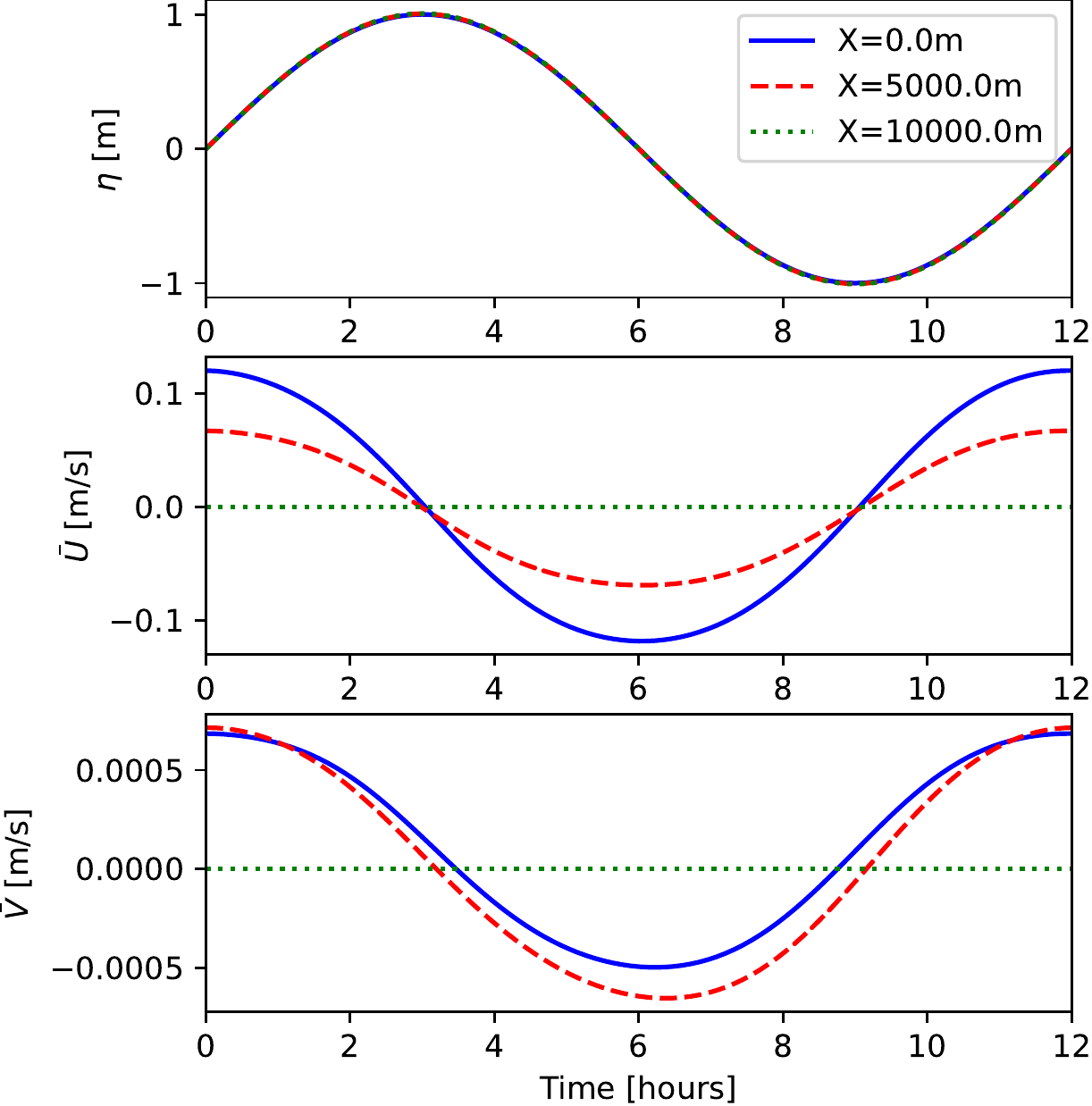}
    \caption{Time series of all three output variables at the beginning, middle and end of the channel. $\eta$ and $\bar{U}$ are measured at the center line, while $\bar{V}$ is measured three quarters across the channel.}
    \label{fig:time_series}
\end{figure}

\begin{figure}[!ht]
    \centering
    \includegraphics[width=0.45\textwidth]{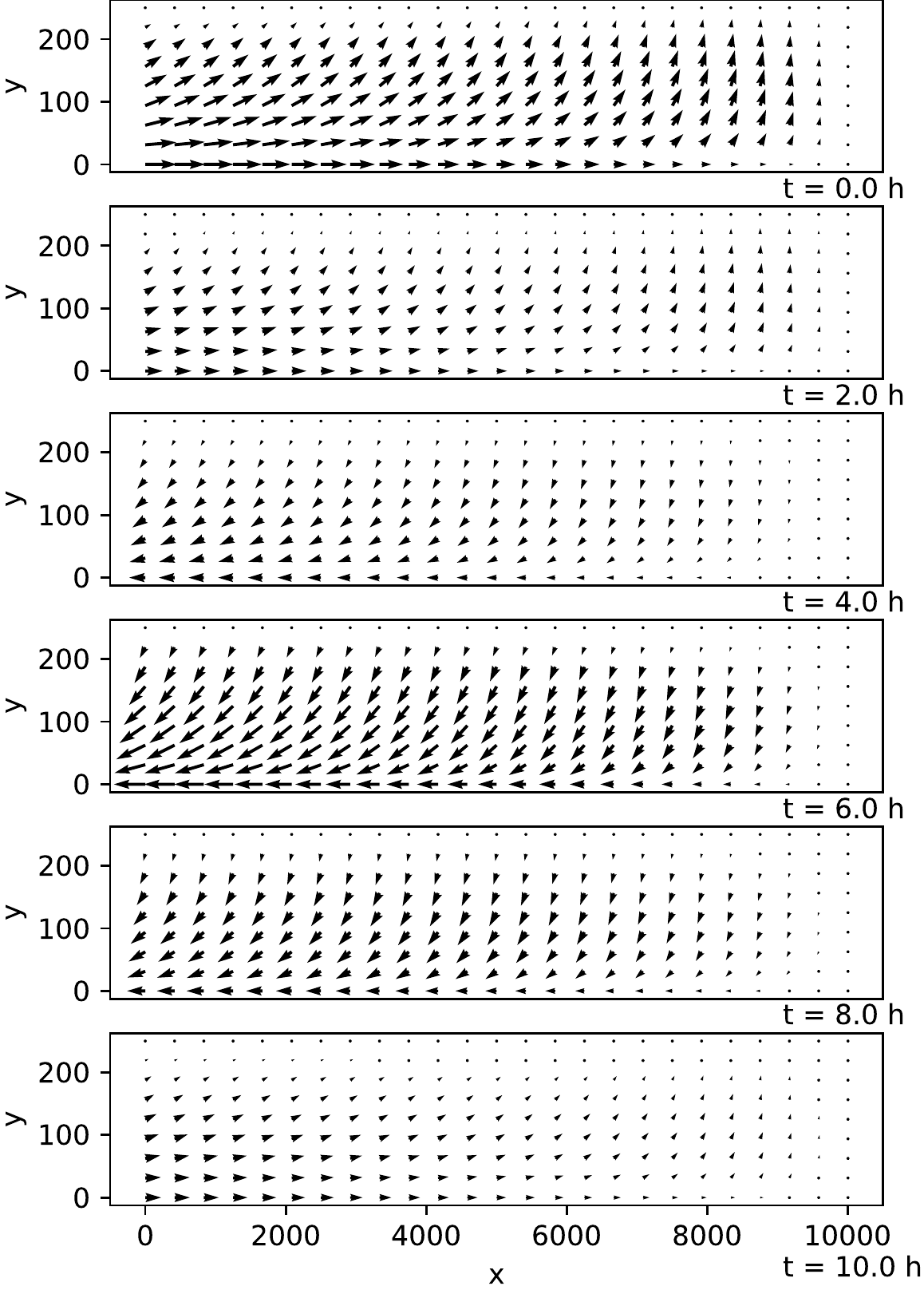}
    \caption{Velocity fields of the solution. One for every 2 hours. The $V$ component was amplified by 100 times for better visualization.}
    \label{fig:velocity_fields}
\end{figure}

Figure~\ref{fig:time_series} shows the time series of all three output variables as predicted by the PINN for the case trained with $n_f=10000$ and resampling, three sampling positions are used, at the beginning, middle and end of the channel; $\eta$ and $\bar{U}$ are measured at the center line, while $\bar{V}$ is measured three quarters across the channel. Figure~\ref{fig:velocity_fields} shows the velocity fields for the same case; one field is shown for every 2 hours of the cycle. Note that the $\bar{V}$ component of the solution was magnified by 100 times for better visualization.

For the performance evaluation of each training batch size and to measure the benefits of resampling the batch after each iteration, we have defined a testing grid consisting of uniformly distributed points along the spatial and temporal domains, where the residual of the governing equations is measured. Due to the non-deterministic nature of the training, we have averaged the residual values from 5 runs of each case. The evaluation grid consists of 101 points in the $x$ direction, 11 points in the $y$ direction and $51$ points in time. We have computed both the mean absolute value (MAE) and the root mean square (RMS) of the residuals.

Additionally, we wish to understand the memory requirements of each batch size; for this, we have used PyTorch's \verb|cuda.max_memory_allocated| method to measure the largest amount of memory allocated by the tensors during training. Along with the total training time, this is an important measure of computational cost because it can be used to estimate the maximum batch size a given system can use. Furthermore, it is important to keep in mind that this case contains a relatively simple geometry and only two spatial dimensions; thus, in case one wishes to study more complex scenarios or include the third spatial dimension in the fluid modeling, it is likely that larger batch sizes will be needed, which leads to the necessity of reducing the memory footprint of the training.

Table~\ref{tab:results} summarizes the results. It is worth noting that the values of $\Delta x$ and $\Delta y$ used in Eq.~\ref{eq:am} for the computation of the Smagorinsky diffusivity were kept consistent with the values used during training. Further in this section, we will discuss the effects of these values on the resulting flow profiles. Comparing the results obtained with and without resampling the training batches, one can see that this technique has a very positive effect on the cases where smaller batches were used but the difference becomes negligible as the batch size increases.

\begin{table*}[]
\centering
\begin{tabular}{c|ccc|cc|cc}
\toprule
\multirow{2}{*}{$n_f$} &  \multirow{2}{*}{Memory (MB)} & Time per & \multirow{2}{*}{GPU usage} & \multicolumn{2}{c|}{No resampling} & \multicolumn{2}{c}{With resampling} \\
       &  & iteration (s) & & RMS            & ABS           & RMS         & ABS          \\
\midrule
300    & 15  & 0.11 & 26\% & 2.5E-4        & 1.3E-4       & 1.2E-4      & 9.7E-5     \\
1000   & 38  & 0.11 & 26\% & 1.7E-4        & 1.1E-4       & 1.1E-4      & 9.2E-5     \\
3000   & 102 & 0.12 & 29\% & 1.1E-4        & 9.2E-5       & 9.4E-5      & 8.8E-5     \\
10000  & 330 & 0.12 & 33\% & 9.5E-5        & 9.5E-5       & 8.8E-5      & 8.9E-5    \\
30000  & 982 & 0.13 & 41\% & 8.6E-5        & 9.2E-5       & 8.8E-5      & 9.1E-5     \\
100000 & 3276 & 0.13 & 88\% & 8.5E-5      & 8.7E-5       & 8.3E-5      & 8.7E-5     \\
\bottomrule
\end{tabular}
\caption{Summary of results for various values of $n_f$.}
\label{tab:results}
\end{table*}

In terms of memory usage, the maximum amount of memory allocated by the tensor's scales almost linearly with the batch size, apart from some overhead. There was no difference in memory allocation whether or not resampling was used. The GeForce\texttrademark~RTX3080 GPU used in these tests has more than enough memory for our largest batch size; however, as discussed earlier, this test scenario is relatively simple and thus may require smaller batch sizes when compared to more complex cases that more closely resemble real-world scenarios; therefore, employing a technique such as this may be crucial when implementing more challenging models, especially for situations where only GPUs with smaller working memories may be available.

The time per iteration for the Adam optimizer was mostly constant for all values of $n_f$. Those cases have also failed to reach 100\% of the GPU usage, as measured by the \verb|nvidia-smi| command. Furthermore, using smaller values of $n_f$ has allowed multiple cases to run simultaneously; for instance, when two cases with $n_f=10000$ were run in parallel, the time per iteration has remained at 0.12 seconds, while the GPU usage has increased to 94\%, up from 33\% for just one case. This illustrates another advantage of smaller batch sizes, i.e., running more than one case at once in the same hardware. In all cases, convergence was achieved after around 50000 iterations, which translates to under 2 hours of real time. For the sake of comparison, the same training was run on a Quadro\texttrademark~P6000 GPU, which is comparatively slower; this time for $n_f\le30000$, each iteration took 0.36 seconds, however, for $n_f=100000$, the GPU reached 100\% usage and the time per iteration was increased to 0.56 seconds.

\begin{figure*}[t]
    \centering
    \includegraphics[width=0.95\textwidth]{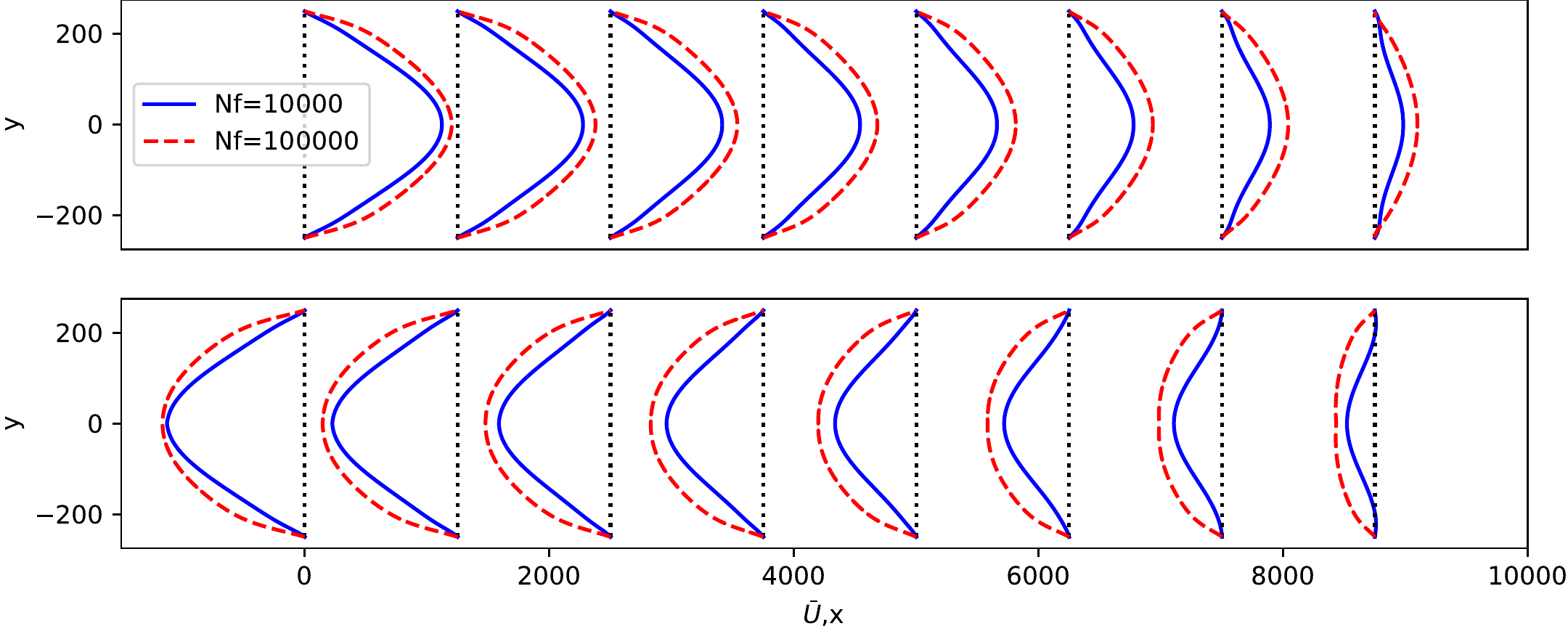}
    \caption{Velocity profiles of the solution for $n_f=10000$ and $n_f=100000$, with resampling. At $t=0 h$ (top) and $t=6h$ (bottom).}
    \label{fig:velocity_profiles}
\end{figure*}

After training, it was noted that by changing the number of function evaluation points, the resulting velocity profile would change considerably, as seen in Fig.~\ref{fig:velocity_profiles}. This is attributed to the computation of the Smagorinsky diffusivity, which considers the grid spacing in each direction to replicate the effect of the sub-grid dissipation observed in more traditional Computational Fluid Mechanics simulations. In the training of this model, we have approximated the grid spacing values as shown in Eq.~\ref{eq:deltax}, similarly to the approximation used in the meshless method of \cite{mramor2022}. In the next section, we discuss some alternatives for the modeling of the sub-grid dissipation in a LES model solved by this type of PINN.

\section{Conclusion}
\label{sec:conclusion}

In this work, we have implemented and trained Physics-Informed Neural Networks to predict the water flow  induced by astronomical tides in a synthetic port channel. With a small addition to the network, in the form of the Fourier features computed for the temporal input, we have shown that a PINN can be trained to take into account the flow periodic in time. There is a considerable advantage in this type of modeling over more traditional forms of fluid simulation. For those forms of simulation, one would have to define an arbitrary initial condition for the flow and then simulate it for several cycles, until the effects of this (usually non-physical) initial condition dissipate and all that is left in the domain is the periodic solution.

For the sake of comparison, we have implemented the same governing equations shown in this paper in a Matlab program that simulates this channel. A regular mesh was used in the domain, consisting of 512 by 32 nodes in the $x$ and $y$ directions, respectively, a sixth-order spectral-like finite differences scheme \cite{lele1992} was implemented for the spatial derivatives, along with a fourth order Runge-Kutta time stepping scheme, with steps of 0.5 seconds. Ten tide cycles were needed to sufficiently dissipate the effects of the initial condition and reach a fully developed flow. This simulation was multi-threaded to 16 CPU cores and took 9 hours to finish. Obviously, it is difficult to compare both scenarios due to the widely different techniques and hardware used. However, this experiment demonstrates the great potential of PINNs for temporally periodic cases, which took under 2 hours to train for the results shown in this paper.

One issue yet to be resolved is the modeling of the sub-grid structures of the LES. \cite{moin2002} argues that instead of relying on the sub-grid dissipation, as done by the POM-rain model, it is best to use a numerical low-pass filter for the LES, which allows better control over the dissipation of small fluid structures. The optimal way of integrating LES modeling with this type of PINN remains an open question. The solution to this issue will require understanding what are the smallest movement scales that each particular neural network can learn. This is the PINN equivalent to the sub-grid scales of a traditional LES model.

One possibility is that the key to modeling the equivalent of a sub-grid resolution of a PINN lies in the eigenspectrum intrinsic to a neural network, as studied by \cite{wang2021}. This eigenspectrum indicates the affinity of a neural network with different spatial and temporal wave numbers in the solution space. Another possibility is to use machine learning models to compute the effect of sub-grid scales \cite{duraisamy2021} -- these models are trained with high-resolution solutions from Direct Numerical Simulations (DNS), that are able to capture the effects of small-scale movements without the help of turbulence modeling due to their very refined grids. \cite{eivazi2021} faced a similar issue when implementing a PINN to solve the Reynolds-Averaged Navier-Stokes equations (RANS). RANS models do not directly solve any eddies in the flow, as opposed to LES which does not solve only the smaller eddies; 
by doing so, the equations are further simplified and the dependence on modeling the sub-grid dissipation is removed, as is the time-dependence. The solution they have found involves adding extra outputs to the neural network to estimate the magnitude of mean velocity fluctuations, which directly relate to the Reynolds stress, which, in turn, is added to the viscous dissipation. Searching for similar solutions for LES models is left as a suggestion for future works.

Finally, we have performed the same training with several batch sizes for the evaluation of the governing equations in the domain and compared models that trained with a fixed batch to those that resampled the training batch after each iteration. For the models that were trained with small batch sizes, resampling the points has considerably increased the model accuracy with little to no added computational cost. This technique may be useful when modeling more complex situations, such as a real channel, with a more intricate geometry, which may require larger batch sizes, or when adding the vertical direction to the governing equations, as this adds a new input dimension and a new output variable to the model.


\section*{Ethical Statement}

There are no ethical issues.

\section*{Acknowledgments}

This work was carried out at the Center for Artificial Intelligence (C4AI-USP), with support by the São Paulo Research Foundation (FAPESP) under grant number 2019/07665-4, and by the IBM Corporation. This work is also supported in part by FAPESP under grant number 2020/16746-5, the Brazilian National Council for Scientific and Technological Development (CNPq) under grant numbers 310085/2020-9, 310127/2020-3, 312180/2018-7, Coordination for the Improvement of Higher Education Personnel (CAPES, Finance Code 001), and by \textit{Ita\'{u} Unibanco S.A.} through the \textit{Programa de Bolsas Ita\'{u}} (PBI) program of the \textit{Centro de Ci\^{e}ncia de Dados} (C$^2$D) of \textit{Escola Polit\'{e}cnica} of USP.

\bibliographystyle{named}
\bibliography{references}

\begin{thebibliography}{}

\bibitem[\protect\citeauthoryear{Beucler \bgroup \em et al.\egroup
  }{2021}]{beucler2021}
Tom Beucler, Michael Pritchard, Stephan Rasp, Jordan Ott, Pierre Baldi, and
  Pierre Gentine.
\newblock Enforcing {{Analytic Constraints}} in {{Neural Networks Emulating
  Physical Systems}}.
\newblock {\em Physical Review Letters}, 126(9), 2021.

\bibitem[\protect\citeauthoryear{Blumberg and Mellor}{1987}]{blumberg1987}
Alan~F. Blumberg and George~L. Mellor.
\newblock A description of a three-dimensional coastal ocean circulation model.
\newblock In N.S. Heaps, editor, {\em Three-{{Dimensional Coastal Ocean
  Models}}}. 1987.

\bibitem[\protect\citeauthoryear{Costa \bgroup \em et al.\egroup
  }{2020}]{costa2020}
Carine~G.R. Costa, Jos{\'e} Roberto~B. Leite, Belmiro~M. Castro, Alan~F.
  Blumberg, Nickitas Georgas, Marcelo Dottori, and Antoni Jordi.
\newblock An operational forecasting system for physical processes in the
  {{Santos-Sao Vicente-Bertioga Estuarine System}}, {{Southeast Brazil}}.
\newblock {\em Ocean Dynamics}, 70(2):257--271, 2020.

\bibitem[\protect\citeauthoryear{Duraisamy}{2021}]{duraisamy2021}
Karthik Duraisamy.
\newblock Perspectives on machine learning-augmented {{Reynolds-averaged}} and
  large eddy simulation models of turbulence.
\newblock {\em Physical Review Fluids}, 6(5):050504, May 2021.

\bibitem[\protect\citeauthoryear{Eivazi \bgroup \em et al.\egroup
  }{2021}]{eivazi2021}
Hamidreza Eivazi, Mojtaba Tahani, Philipp Schlatter, and Ricardo Vinuesa.
\newblock Physics-informed neural networks for solving {{Reynolds-averaged
  Navier-Stokes}} equations, July 2021.

\bibitem[\protect\citeauthoryear{Fukami \bgroup \em et al.\egroup
  }{2020}]{fukami2020}
Kai Fukami, Koji Fukagata, and Kunihiko Taira.
\newblock Machine-learning-based spatio-temporal super resolution
  reconstruction of turbulent flows.
\newblock {\em Journal of Fluid Mechanics}, 909, 2020.

\bibitem[\protect\citeauthoryear{He \bgroup \em et al.\egroup }{2016}]{he2016}
Kaiming He, Xiangyu Zhang, Shaoqing Ren, and Jian Sun.
\newblock Deep {{Residual Learning}} for {{Image Recognition}}.
\newblock In {\em 2016 {{IEEE Conference}} on {{Computer Vision}} and {{Pattern
  Recognition}} ({{CVPR}})}, pages 770--778, June 2016.

\bibitem[\protect\citeauthoryear{Karniadakis \bgroup \em et al.\egroup
  }{2021}]{karniadakis2021}
George~Em Karniadakis, Ioannis~G. Kevrekidis, Lu~Lu, Paris Perdikaris, Sifan
  Wang, and Liu Yang.
\newblock Physics-informed machine learning.
\newblock {\em Nature Reviews Physics}, 3(6):422--440, 2021.

\bibitem[\protect\citeauthoryear{Kashinath \bgroup \em et al.\egroup
  }{2021}]{kashinath2021}
K.~Kashinath, M.~Mustafa, A.~Albert, J.~L. Wu, C.~Jiang, S.~Esmaeilzadeh,
  K.~Azizzadenesheli, R.~Wang, A.~Chattopadhyay, A.~Singh, A.~Manepalli,
  D.~Chirila, R.~Yu, R.~Walters, B.~White, H.~Xiao, H.~A. Tchelepi, P.~Marcus,
  A.~Anandkumar, P.~Hassanzadeh, and {Prabhat}.
\newblock Physics-informed machine learning: {{Case}} studies for weather and
  climate modelling.
\newblock {\em Philosophical Transactions of the Royal Society A: Mathematical,
  Physical and Engineering Sciences}, 379(2194), 2021.

\bibitem[\protect\citeauthoryear{Kolmogorov}{1991}]{kolmogorov1991}
A.~N. Kolmogorov.
\newblock The {{Local Structure}} of {{Turbulence}} in {{Incompressible Viscous
  Fluid}} for {{Very Large Reynolds Numbers}}, 1991.

\bibitem[\protect\citeauthoryear{Lele}{1992}]{lele1992}
Sanjiva~K. Lele.
\newblock Compact finite difference schemes with spectral-like resolution.
\newblock {\em Journal of Computational Physics}, 103(1):16--42, November 1992.

\bibitem[\protect\citeauthoryear{Li \bgroup \em et al.\egroup }{2021}]{li2021}
Zongyi Li, Nikola~Borislavov Kovachki, Kamyar Azizzadenesheli, Burigede Liu,
  Kaushik Bhattacharya, Andrew Stuart, and Anima Anandkumar.
\newblock Fourier {{Neural Operator}} for {{Parametric Partial Differential
  Equations}}.
\newblock In {\em International {{Conference}} on {{Learning
  Representations}}}, 2021.

\bibitem[\protect\citeauthoryear{Moin}{2002}]{moin2002}
Parviz Moin.
\newblock Advances in large eddy simulation methodology for complex flows.
\newblock {\em International Journal of Heat and Fluid Flow}, 23(5):710--720,
  October 2002.

\bibitem[\protect\citeauthoryear{Mramor \bgroup \em et al.\egroup
  }{2022}]{mramor2022}
Katarina Mramor, Robert Vertnik, and Bo{\v z}idar {\v S}arler.
\newblock Meshless approach to the large-eddy simulation of the continuous
  casting process.
\newblock {\em Engineering Analysis with Boundary Elements}, 138:319--338, May
  2022.

\bibitem[\protect\citeauthoryear{Nair \bgroup \em et al.\egroup
  }{2019}]{nair2019}
Aditya~G. Nair, Chi~An Yeh, Eurika Kaiser, Bernd~R. Noack, Steven~L. Brunton,
  and Kunihiko Taira.
\newblock Cluster-based feedback control of turbulent post-stall separated
  flows.
\newblock {\em Journal of Fluid Mechanics}, 875(M):345--375, 2019.

\bibitem[\protect\citeauthoryear{Raissi \bgroup \em et al.\egroup
  }{2019}]{raissi2019}
M.~Raissi, P.~Perdikaris, and G.~E. Karniadakis.
\newblock Physics-informed neural networks: {{A}} deep learning framework for
  solving forward and inverse problems involving nonlinear partial differential
  equations.
\newblock {\em Journal of Computational Physics}, 378:686--707, 2019.

\bibitem[\protect\citeauthoryear{Read \bgroup \em et al.\egroup
  }{2019}]{read2019}
Jordan~S. Read, Xiaowei Jia, Jared Willard, Alison~P. Appling, Jacob~A. Zwart,
  Samantha~K. Oliver, Anuj Karpatne, Gretchen~J.A. Hansen, Paul~C. Hanson,
  William Watkins, Michael Steinbach, and Vipin Kumar.
\newblock Process-{{Guided Deep Learning Predictions}} of {{Lake Water
  Temperature}}.
\newblock {\em Water Resources Research}, 55(11):9173--9190, 2019.

\bibitem[\protect\citeauthoryear{Sun \bgroup \em et al.\egroup
  }{2020}]{sun2020}
Luning Sun, Han Gao, Shaowu Pan, and Jian-Xun Wang.
\newblock Surrogate modeling for fluid flows based on physics-constrained deep
  learning without simulation data.
\newblock {\em Computer Methods in Applied Mechanics and Engineering},
  361:112732, April 2020.

\bibitem[\protect\citeauthoryear{Wang \bgroup \em et al.\egroup
  }{2021}]{wang2021}
Sifan Wang, Hanwen Wang, and Paris Perdikaris.
\newblock On the eigenvector bias of {{Fourier}} feature networks: {{From}}
  regression to solving multi-scale {{PDEs}} with physics-informed neural
  networks.
\newblock {\em Computer Methods in Applied Mechanics and Engineering},
  384:113938, October 2021.

\bibitem[\protect\citeauthoryear{Willard \bgroup \em et al.\egroup
  }{2020}]{willard2020}
Jared Willard, Xiaowei Jia, Shaoming Xu, Michael Steinbach, and Vipin Kumar.
\newblock Integrating {{Scientific Knowledge}} with {{Machine Learning}} for
  {{Engineering}} and {{Environmental Systems}}.
\newblock 1(1):1--34, March 2020.

\bibitem[\protect\citeauthoryear{Wu \bgroup \em et al.\egroup }{2020}]{wu2020}
Mengning Wu, Christos Stefanakos, and Zhen Gao.
\newblock Multi-step-ahead forecasting of wave conditions based on a
  physics-based machine learning ({{PBML}}) model for marine operations.
\newblock {\em Journal of Marine Science and Engineering}, 8(12):1--24, 2020.

\bibitem[\protect\citeauthoryear{Xu and Valocchi}{2015}]{xu2015}
Tianfang Xu and Albert~J. Valocchi.
\newblock Data-driven methods to improve baseflow prediction of a regional
  groundwater model.
\newblock {\em Computers and Geosciences}, 85:124--136, 2015.

\bibitem[\protect\citeauthoryear{Zhiyin}{2015}]{zhiyin2015}
Yang Zhiyin.
\newblock Large-eddy simulation: {{Past}}, present and the future.
\newblock {\em Chinese Journal of Aeronautics}, 28(1):11--24, February 2015.

\end{thebibliography}

\end{document}